%% file: NEQI_Screening.tex
\documentclass[twocolumn,superscriptaddress]{revtex4}
\usepackage{dcolumn}
\usepackage{graphicx}
\usepackage{latexsym}
\usepackage{amsfonts}
\usepackage{amssymb}
\usepackage{dsfont}

\usepackage{amssymb,amsmath,stmaryrd}
\usepackage{times}
\usepackage{float}
\usepackage{amsmath}
\usepackage{pst-node}
\usepackage[nouppercase]{scrpage2}
\usepackage{makeidx}
\usepackage{bbm}
\usepackage{cancel}
\usepackage{amsthm}
\usepackage{textgreek}

\makeatletter
\newcommand{\subalign}[1]{%
  \vcenter{%
    \Let@ \restore@math@cr \default@tag
    \baselineskip\fontdimen10 \scriptfont\tw@
    \advance\baselineskip\fontdimen12 \scriptfont\tw@
    \lineskip\thr@@\fontdimen8 \scriptfont\thr@@
    \lineskiplimit\lineskip
    \ialign{\hfil$\m@th\scriptstyle##$&$\m@th\scriptstyle{}##$\crcr
      #1\crcr
    }%
  }
}
\makeatother

\DeclareGraphicsExtensions{.pdf,.gif,.jpg}

\input{ourmacros}

\begin{document}

% \title{Pump-induced topological excitonic insulator: Floquet Majorana 
% fermions Josephson oscillations }
\title{Screening properties of the nonequilibrium excitonic insulator}

\title{Self-consistent screening enhances stability of the nonequilibrium excitonic 
insulator phase}

\author{E. Perfetto}
\affiliation{Dipartimento di Fisica, Universit\`{a} di Roma Tor Vergata,
Via della Ricerca Scientifica 1, 00133 Rome, Italy}

\author{A. Marini}
\affiliation{Istituto di Struttura della Materia of the National Research 
Council, Via Salaria Km 29.3, I-00016 Montelibretti, Italy}

\author{G. Stefanucci}
\affiliation{Dipartimento di Fisica, Universit\`{a} di Roma Tor Vergata,
Via della Ricerca Scientifica 1, 00133 Rome, Italy}
\affiliation{INFN, Laboratori Nazionali di Frascati, Via E. Fermi 40, 00044 Frascati, 
Italy}

\begin{abstract}
The nonequilibrium excitonic insulator (NEQ-EI) is an excited state of matter 
characterized by a finite density of coherent excitons and a  time-dependent  
macroscopic polarization. The stability of this exciton superfluid as the density 
grows is jeopardized by the increased screening efficiency of the 
looser excitons. In this work  we  put forward a 
Hartree plus Screened Exchange HSEX scheme to predict the critical 
density at which the transition toward a 
free electron-hole plasma occurs. 
The dielectric function is calculated self-consistently using 
the NEQ-EI polarization and found to vanish in the long-wavelength limit.
This property makes the exciton superfluid 
stable up to relatively high densities. Numerical results for the  MoS$_{2}$ 
monolayers indicate that the NEQ-EI phase survives up to 
densities of the order of $10^{12}\mathrm{cm}^{-2}$. 
\end{abstract}

\maketitle

\section{Introduction}

The significant experimental activity in exploring 
% impressive experimental findings in 
atomically-thin transition 
metal dichalcogenides (TMD)~\cite{tmd1,tmd3,tmd4,tmd5,tmd7}  
has renewed the interest and boosted the research
on the physics of excitons. 
Optically excited TMD are indeed characterized 
by quasi-free carriers and, due to the relatively strong Coulomb 
interaction~\cite{exc3,exc5,tmd6}, by a rich manifold of excitonic states 
like  bound excitons, charged excitons 
(trions)~\cite{trion1,trion2,trion3}, excitonic 
molecules (biexcitons)~\cite{biex1,biex2,biex3,biex4} as well as
exciton-polariton complexes~\cite{expol1,expol2,expol3}.
% because of the remarkably strong Coulomb 
% interaction that chracterizes these materials.~\cite{exc3,exc5,tmd6}
% Here optical excitations create 
% complex excited states with a finite carrier population  
% in the conduction band. This population can 
% be attribuited  to quasi-free-carriers and to a rich manyfold of excitonic states 
% like  bound excitons, charged excitons 
% (trions)~\cite{trion1,trion2,trion3}, excitonic 
% molecules (biexcitons)~\cite{biex1,biex2,biex3,biex4}, 
% exciton-polariton complexes~\cite{expol1,expol2,expol3} etc. 
Excitons do therefore play 
% For this reason excitons play 
a prominent role in determining  optical and electronic 
properties and leave clear fingerprints in 
% of these photo-excited semiconductors, as pronounced 
% resonances originating from different excitonic states 
% dominate both 
photoabsorption and photoluminescence
spectra~\cite{exc3,exc5,exc1,exc2,exc4,exc6}. 
Establishing the amount of excitable excitons and the nature of the 
exciton fluid are among the most 
interesting and investigated issues.

The rich excitonic phenomenology in complex materials can be 
efficiently investigated using pump\&probe techniques. A first 
laser pulse (pump) excites the material which is subsequently probed 
by a second, weaker pulse sent with a tunable delay from the pump. 
Depending on the pump-probe  delay an incoherent 
and a coherent  regime can be identified. 
% One immediately after the pump is 
% largely dominated by the coherence induced by the pump field. The 
% second regime oc- curs far from the pump where coherence has been 
% destroied by the internal collisional processes of the 
% material.          
At delays of the order of tens of picoseconds, 
coherence is destroyed by 
carrier-carrier~\cite{carcar0,carcar} and 
carrier-phonon~\cite{carphon,carphon2} scattering processes. The 
system reaches a quasi-equilibrium state characterized by 
quasi-free carriers coexisting with  
{\em incoherent} excitons~\cite{scattering,explasma1,psms2016}.
The quasi-free carriers efficiently screen the electron-hole 
attraction thus reducing both the exciton 
binding energy~\cite{exweak1,exweak2,exweak3} and  the bandgap~\cite{exweak2,exweak3,bgr1,bgr2,bgr3}.
% Screening  is also responsible for renormalizing 
% the bandgap~\cite{exweak2,exweak3,bgr1,bgr2,bgr3}. 
For large enough 
density  of quasi-free carriers the exciton binding energy  becomes 
comparable with the bandgap and excitons 
ionize~\cite{mott5,bgr2,exweak1,dendzik}, a phenomenon called 
excitonic Mott transition~\cite{mott1,mott2,mott3,mott4}. 
% However, the total amount of created excitons is limited by many-body effects
% between their constituents (electrons and holes).
% Indeed in the excited state the mobility of the elementary excitations
% is much higher than that in 
% the ground-state, leading to a substancial enhancement of the screening 
% efficiency.
% As a consequence, by increasing the carrier density $n$, 
% the effective Coulomb attraction between electrons and holes weakens,
% leading to a reduction of the exciton 
% binding.~\cite{exweak1,exweak2,exweak3}
% Screening in the excited state is also responsible for 
% determining the magnitude of the bandgap renormalization in the 
% material.~\cite{exweak2,exweak3,bgr1,bgr2,bgr3} 
% When $n$
% exceeds a critical value, the bandgap narrowing induced by the screened
% interaction can lower the conduction band minimum below the energy of the 
% original exciton peak, thus causing the exciton 
% ionization.~\cite{mott5,bgr2,exweak1,dendzik} This 
% phenomenon is known as excitonic Mott 
% transition.~\cite{mott1,mott2,mott3,mott4}
The simplest approach to estimate the screened interaction in this 
incoherent regime consists 
in 
% From the theoretical side it is clear that a careful evaluation 
% of the effective screened interaction in the excited state is crucial for a 
% quantitative estimate of the various excitonic effects.
% The simplest approach consists in 
evaluating the dielectric function assuming that {\it all}  
excited carries are free~\cite{plasma1,plasma2,plasma3}.
The RPA approximation yields a 
plasma-screened Coulomb interaction that in TMD monolayers
leads to a strong bandgap shrinkage 
and a sizable reduction of the exciton binding energy even at
moderate densities~\cite{plasma1}.
%, as experimentally observed~\cite{bgr2,exweak1}.
However, excited carriers partially form bound 
excitons which are neutral composite excitations and hence have a 
scarce screening efficiency.
It is therefore important to balance  free-carrier 
versus exciton
contributions in the dielectric function~\cite{excscr1,excscr2,excscr3}.
% This task is very challenging since the screening properties of excitons 
% strongly depend on the nature of the many-body excited state. 
% At long times (of the order of tens of picoseconds) after pumping, incoherent effects caused by 
% carrier-carrier~\cite{carcar0,carcar} and 
% carrier-phonon~\cite{carphon,carphon2} scattering drive the system 
% towards a relaxed and dephased regime characterized by a quasi-equilibrium between plasma 
% and excitons.~\cite{scattering,explasma1,psms2016}
% Here the total 
% dielectric function consists in the sum of a Lindhard-like function and an 
% exciton part, written in terms of an equilibrium Bose distribution
% and excitonic energies and 
% wavefunctions.~\cite{excscr1,excscr2,excscr3}
Approaches in this direction~\cite{excscr4} indicate that
% has been worked out in 
% Ref.~\cite{excscr4} showing that 
% Using this more refined strategy it 
% it is possible to address 
% quantitatively the stability of the exciton gas
% with respect to $n$ and temperature. It 
% has been shown~\cite{excscr4} that 
a phase dominated by excitons in TMD monolayers  
can survive up to relatively high densities 
$n \sim 10^{13}\mathrm{cm^{-2}}$, consistently with 
the experimental data~\cite{mott5,bgr2,exweak1}.
%      at low temperature, in thermal equilibrium for example,
%      excitons can easily outnumber free carriers. As a result, 
%      the screening of the electron-hole interaction due to excitons,
%      which is in principle weak,
%      may become important, and even stronger than the screening due to free carriers.
%      The stability of the exciton gas
%      versus the density and the temperature in the static approximation of screening. 
%      A special feature is the weakening of the coupling with increasing
%      density caused by many-particle effects such as screening of the Coulomb
%      interaction and, therefore, a breakup of the excitons usually referred to
%      as Mott effect density ionization.3-6
%      The Mott effect is, in principle, understood as caused by lowering of the band edge,
%      (“band-gap  renormalization”  (BGR)  effect)
%      while the exciton energy only weakly changes with increasing density.
%      plane. The Mott effect occurs roughly at rsc $=$ aX, where rsc
%      denotes the screening length and ax the excitonic Bohr radius.

The coherent regime does instead set in immediately after the pump 
and survives until scattering induced dephasing mechanism destroy the 
coherence brought by the laser.
It has been predicted in a number of 
papers that a {\em coherent} exciton fluid, or exciton superfluid, 
can be realized by pumping resonant with the exciton absorption 
peak of a normal semiconductor (or 
insulator)~\cite{neqei1,neqei10,neqei11,neqei12,neqei13,neqei14,neqei2,neqei3,neqei4,neqei5,neqei6,neqei7,neqei8,neqei9}.
Experimental evidence has been recently reported 
in GaAs by optical pump-probe spectroscopy~\cite{murotani}. 
% Just after pumping and before any scattering induced 
% dephasing mechanism becomes effective  the  system is temporarily left in a 
% NEQ-EI phase~\cite{neqei1,neqei10,neqei11,neqei12,neqei13,neqei14,neqei2,neqei3,neqei4,neqei5,neqei6,neqei7,neqei8,neqei9}.
We stress here that the
superfluid phase is not exclusive of excited states as it can 
be found in the ground state too.
% can be realized also in the ground state or in some higher energy state of the material.
The system is said to be an Excitonic 
Insulator (EI) in the latter case and a nonequilibrium (NEQ) EI in 
the former case. Exciton superfluids are 
characterized by a finite exciton
population and by a steady (EI) or oscillatory  (NEQ-EI) 
macroscopic polarization. 
The EI phase of semimetals and small gap semiconductors
has been proposed long 
ago~\cite{eqei1,eqei2,eqei3,eqei4,eqei5,eqei6,eqei7}. 
Calculations on the stability of the EI phase against screening effects 
have been pioneered by Nozieres and Compte~\cite{noz}, and subsequently
performed in different bilayered compounds, including dipolar 
systems~\cite{conscrbily}, 
graphene~\cite{conscrgraph1,conscrgraph2,conscrgraph3}, 
and TMD~\cite{conscrtmd}. However, how a screened electron-hole interaction 
affects the stability of a 
NEQ-EI has, to our knowledge, not yet been addressed.
It is the purpose of this work to contribute in filling the gap.

% The effects of a screened interaction in the coherent regime 
% have not received the same attention as in the incoherent regime. 

% It has been predicted by a number of 
% papers that NEQ-EI's can be realized by pumping resonant with the exciton absorption 
% peak of a normal semiconductor (or 
% insulator)~\cite{neqei1,neqei10,neqei11,neqei12,neqei13,neqei14}.
% Just after pumping and before any scattering induced 
% dephasing mechanism becomes effective  the  system is temporarily left in a 
% NEQ-EI phase~\cite{neqei1,neqei10,neqei11,neqei12,neqei13,neqei14,neqei2,neqei3,neqei4,neqei5,neqei6,neqei7,neqei8,neqei9}.
% This phase has been 
% recently observed in GaAs by optical pump-probe 
% spectroscopy~\cite{murotani}. 
% However, theoretical studies up to date have 
% ignored screening effects.
The difficulty in addressing screening effects in NEQ-EI is two-fold: 
the system is neither in equilibrium nor in a stationary state since
the macroscopic polarization features self-sustained (monochromatic) oscillations. 
% In such  broken-symmetry state the excitons 
% condense and, depending on the excited density $n$, they can form a 
% Bose-Einstein condensate or a BCS superfluid.~\cite{neqei11,neqei13} 
% Also in this case, however, the larger the excited density is, the 
% stronger the generated screening results, and 
% therefore exciton desruption is expected at sufficiently large 
% $n$.~\cite{semkat} 
In this work we  put forward a {\em self-consistent} Hartree
plus Screened Exchange (HSEX) {\em nonequilibrium} scheme 
which overcomes the aforementioned difficulties and 
allows us to assess quantitatively the role 
of  screening in an 
exciton superfluid.
% Here we extend these studies to out-of-equilibrium {\it excited states}
% orginating from a photoexcited normal semiconductor. We address stability 
% of the resulting NEQ-EI state against the screening self-generated
% by the excitonic condensate. 
Unlike the dielectric function in the incoherent regime  we find that 
the dielectric function of a NEQ-EI cannot be written as the sum of a plasmonic and 
excitonic contributions since the two
are intimately entangled. 
We also show that the long-wavelength component of the dielectric function
vanishes, making the NEQ-EI phase particularly 
robust.
Numerical evidence  is provided for MoS$_{2}$ 
monolayers where the NEQ-EI phase is predicted to survive up to 
$n\sim  10^{12}\mathrm{cm}^{-2}$. 

% We recall that the occurrence of an EI phase in the {\it ground-state}
% of semimetals and small gap semiconductors
% has been proposed long 
% ago.~\cite{eqei1,eqei2,eqei3,eqei4,eqei5,eqei6,eqei7} 
% The screening properties of the corresponding exciton superfluid
% have been pioneered by Nozieres and Compte~\cite{noz}, and subsequently
% computed in different bilayered compounds, including dipolar 
% systems,~\cite{conscrbily} 
% graphene,~\cite{conscrgraph1,conscrgraph2,conscrgraph3} 
% and TMD.~\cite{conscrtmd} 
% Here we extend these studies to out-of-equilibrium {\it excited states}
% orginating from a photoexcited normal semiconductor. We address stability 
% of the resulting NEQ-EI state against the screening self-generated
% by the excitonic condensate. In this case 
% the total dielectric function cannot be written simply as sum of a plasma and 
% an excitonic part, as the two contributions are intimately entangled.
% Nevertheless it can be shown that as long as the system hosts a 
% finite order parameter the long-wave component of the dielectric function
% vanishes. This remarkable property renders the NEQ-EI particularly 
% robust, at least for small and moderate excited density $n$.
% Numerical evidence for the stability of the  NEQ-EI phase is provided for MoS$_{2}$ 
% monolayers, where the excitonic condensate is predicted to survive up to 
% $n\sim  10^{12}\mathrm{cm}^{-2} $. 

The paper is organized as follows. In Section~\ref{sec1} we introduce
the model Hamiltonian for a two-band semiconductor, and we briefly review the
Hartree-Fock (HF) theory of the NEQ-EI phase.
In Section~\ref{screensec} we  calculate the 
polarization function of the exciton superfluid and use it 
to screen the electron-hole interaction at the RPA level.
In Section~\ref{sec3} we improve over the HF results by laying down a 
self-consistent HSEX theory which we solve numerically. 
Results for the phase diagram in monolayer 
MoS$_{2}$ are discussed in Section~\ref{mos2}.
A summary and the main conclusions are drawn in 
Section~\ref{concl}.

\section{Hartree-Fock NEQ-EI}
\label{sec1}

We consider a semiconductor (or insulator) with 
one valence band of bare dispersion $\e^{\mathrm{b}}_{v \blk }$  and one conduction 
band of bare dispersion $\e^{\mathrm{b}}_{c \blk }$.
The explicit form of the Hamiltonian reads
\bea
\hat{H}&=&\sum_{\blk \s }(\e^{\mathrm{b}}_{v \blk 
}\hat{v}^{\dag}_{\blk \s}\hat{v}_{\blk \s}
+\e^{\mathrm{b}}_{c \blk }\hat{c}^{\dag}_{\blk \s}\hat{c}_{\blk \s})
\nn\\
&+&
\frac{1}{2\callN}\sum_{\blk_{1}\blk_{2}\blq \s 
\s'}U^{\blq}_{ vv}\,\hat{v}^{\dag}_{\blk_{1}+\blq 
\s}\hat{v}^{\dag}_{\blk_{2}-\blq \s'}
\hat{v}_{\blk_{2} \s'}\hat{v}_{\blk_{1} \s} \nn\\
&+&
\frac{1}{2\callN}\sum_{\blk_{1}\blk_{2}\blq \s 
\s'}U^{\blq}_{cc}\,\hat{c}^{\dag}_{\blk_{1}+\blq 
\s}\hat{c}^{\dag}_{\blk_{2}-\blq \s'}
\hat{c}_{\blk_{2} \s'}\hat{c}_{\blk_{1} \s} \nn\\
&+&
\frac{1}{\callN}\sum_{\blk_{1}\blk_{2}\blq \s 
\s'}U^{\blq}_{cv}\,\hat{v}^{\dag}_{\blk_{1}+\blq 
\s}\hat{c}^{\dag}_{\blk_{2}-\blq \s'}
\hat{c}_{\blk_{2} \s'}\hat{v}_{\blk_{1}\s},
\label{minmodham}
\eea
where
$\hat{v}_{\blk \s}$ ($\hat{c}_{\blk \s}$) annihilates an electron of 
momentum $\blk$ and spin $\s$ in the 
valence (conduction) band, $U^{\blq}_{\m\n}=U^{\blq}_{\n\m}$ is the (spin-independent)  Coulomb 
interaction between electrons in bands $\m$ and $\n$, and $\callN$ is 
the number of discretized $\blk$-points.
In Eq.~(\ref{minmodham}) we have assumed that the interaction preserves the number of 
particles 
in each band since Coulomb integrals that break this property are tipically  
small~\cite{smallv}.
All derivations below can be 
easily generalized to the case of multiple bands.

In this section we review the unscreened HF characterization of the NEQ-EI state. 
According to Ref.~\onlinecite{neqei11} the NEQ-EI state can be 
found by solving a self-consistent eigenvalue problem characterized 
by {\em  different} chemical potentials $\m_{v}$ and $\m_{c}$ for valence and conduction 
electrons respectively. In a $2\times 2$ matrix form the 
self-consistent equations read
\be
\left[h_{\blk}+V^{\rm HF}_{\blk}-\mu+\frac{\d\m}{2}\s_{z}\right]\vec{\vf}^{\l}_{\blk}=
e^{\l}_{\blk}\vec{\vf}^{\l}_{\blk},\quad \l=\pm
\label{scprho}
\ee
where we have defined the center-of-mass chemical potential $\m=\frac{\m_{v}+\m_{c}}{2}$, 
the relative chemical potential $\d\m=\m_{c}-\m_{v}$ and
the bare single particle Hamiltonian with 
matrix elements
$h_{\blk}^{\m\n}=\d_{\m\n}\e^{\mathrm{b}}_{\m \blk }$. The HF 
potential $V^{\rm HF}_{\blk}$ in Eq.~(\ref{scprho}) is the following functional of the 
one-particle density matrix $\r^{\m \n}_{\blk \s \s'}=\d_{\s \s'}\r^{\m 
\n}_{\blk}$
\bea
V^{{\rm HF}, vv}_{\blk}&=&\frac{1}{\callN}
\sum_{\blq} \left(2U^{\mathbf{0}}_{vv}
\r^{vv}_{\blq}   + 2U^{\mathbf{0}}_{cv}\r^{cc}_{\blq} -  
U^{\blq}_{vv}  \r^{vv}_{\blk-\blq} \right),  \nn \\
V^{{\rm HF},cc}_{\blk}&=&\frac{1}{\callN}
\sum_{\blq} \left(2U^{\mathbf{0}}_{cc}\r^{cc}_{\blq} +
2U^{\mathbf{0}}_{cv} \r^{vv}_{\blq} -  
U^{\blq}_{cc} \r^{cc}_{\blk-\blq}  \right), \nn \\
V^{{\rm HF},cv}_{\blk}&=&V^{{\rm HF},vc}_{\blk}=  -  
\frac{1}{\callN} \sum_{\blq} U^{\blk-\blq}_{cv} \r^{cv}_{\blq}.
\label{hfpot}
\eea
The self-consistency emerges when expressing the density 
matrix in terms of  the eigenvectors:
\be
\r^{\m\n}_{\blk}=\sum_{\l}f(e^{\l}_{\blk})\vf^{\l}_{\m\blk}\vf^{\l\ast}_{\n\blk},
\label{rhowitheigvect}
\ee
where $f$ is the Fermi function.
In equilibrium $\d\m=0$ and at zero temperature the 
chemical potential $\m$ is such that $\r_{\blk}=\r^{{\rm gs}}_{\blk}$
with $\r^{{\rm gs},vv}_{\blk}=1$ and 
$\r^{{\rm gs},cc}_{\blk}=\r^{{\rm gs},cv}_{\blk}=0$
(filled valence band and empty conduction 
band). It is straighforward to verify that in this case 
$h_{\blk}+V^{\rm HF}_{\blk}$ is a diagonal $2\times 2$ matrix with 
diagonal elements
\bea
\e^{\mathrm{HF}}_{v \blk } &=& \e^{\mathrm{b}}_{v \blk } + 
2U^{\mathbf{0}}_{vv} -\frac{1}{\callN}\sum_{\blq}U^{\blq}_{vv}  ,
  \nn \\
\e^{\mathrm{HF}}_{c \blk } &=& \e^{\mathrm{b}}_{v \blk } + 
2U^{\mathbf{0}}_{vc}  .
\label{hfshift}
\eea
Excited states solution are obtained for $\d\m\neq 0$. For 
these solutions to have the same number of electrons as in the ground 
state (charge neutrality condition) the chemical potential $\m$ must be chosen in such a way that 
\be
N_{\rm 
el}=2\sum_{\blk}\Tr[\r_{\blk}]=2\sum_{\l}\sum_{\blk}f(e^{\l}_{\blk})
=2\sum_{\blk}\r^{{\rm gs},vv}_{\blk}.
\ee

Without loss of generality we assume real Coulomb integrals 
$U^{\m \n}_{\blq}$ and choose the normalized eigenvectors 
$\varphi^{\l}_{\m \blk}$ as real vectors.
Let us  cast the self-consistent problem in a slightly different 
form. We write
\be
\r_{\blk}=\r^{\rm gs}_{\blk}+\d\r_{\blk}.
\ee
Then Eq.~(\ref{scprho}) is transformed into a self-consistent 
equation for the variation $\d\r_{\blk}$:
\be
\left[h_{\blk}^{\rm HF}+\d V^{\rm HF}_{\blk}-\frac{\d\m}{2}\s_{z}\right]\vec{\vf}^{\l}_{\blk}=
(e^{\l}_{\blk}+\m)\vec{\vf}^{\l}_{\blk},\quad \l=\pm
\label{scpdeltarho}
\ee
where  $\d V^{\rm HF}_{\blk}$ is defined as in Eqs.~(\ref{hfpot}) with 
$\r_{\blk}\to\d\r_{\blk}$ and $h_{\blk}^{{\rm HF},\m\n}=\d_{\m\n}\e^{\rm HF}_{\m}$. 
Using Eq.~(\ref{rhowitheigvect}) we can easily express 
$\d\r_{\blk}$ in terms of the eigenvectors
\be
\d\r_{\blk}^{\m\n}=\sum_{\l}f(e^{\l}_{\blk})\vf^{\l}_{\m\blk}\vf^{\l}_{\n\blk}-
\r^{\rm gs,\m\n}_{\blk},
\label{drwf}
\ee
while the condition of charge neutrality becomes
\be
\sum_{\blk}\Tr[\d\r_{\blk}]=\sum_{\blk}\left[\sum_{\l}f(e^{\l}_{\blk})-1\right]=0.
\label{cncdr}
\ee
We assume that the band structure is regular enough 
for guaranteeing the existence of a chemical potential such that 
${\rm max}_{\blk}\{e^{-}_{\blk}\}<{\rm 
min}\{e^{+}_{\blk}\}$. Then $f(e^{-}_{\blk})=1$ and 
$f(e^{+}_{\blk})=0$ for all $\blk$ and
Eq.~(\ref{cncdr}) is automatically satisfied. Furthermore, 
Eq.~(\ref{drwf}) implies
\bea
 \d \r^{cc}_{\blk}&=&-\d \r^{vv}_{\blk} = (\varphi^{-}_{c \blk})^{2} ,
 \nn \\
 \d \r^{cv}_{\blk}&=&\varphi^{-}_{c \blk} \varphi^{-}_{v \blk}  .
\eea

In Ref.~\onlinecite{neqei11} we have shown that if the difference 
between the chemical potentials is larger than the lowest exciton energy 
$\e_{\mathrm{x}}$, i.e.,  $\d \m = 
\m_{c}-\m_{v}>\e_{\mathrm{x}}$, then the self-consistent problem in 
Eq.~(\ref{scpdeltarho}) admits a NEQ-EI solution. It is characterized by
a spontaneous symmetry breaking with finite order parameter
\be
\D\equiv \d V^{{\rm HF},cv}_{\blk=\mathbf{0}}.
\label{orpar}
\ee
In the next sections we discuss the robustness of the HF NEQ-EI phase 
against the screening of electrons in the excited state. In fact, the HF NEQ-EI
solution is characterized by a {\it finite} density of electrons in 
the conduction band
that, in principle, could lead to a sizable reduction of the 
Coulomb electron-hole attraction 
and, therefore, to the restoration of a symmetry unbroken phase.

\section{Screened interaction in the NEQ-EI phase}
\label{screensec}

\begin{figure}[t]
\includegraphics[width=0.4\textwidth]{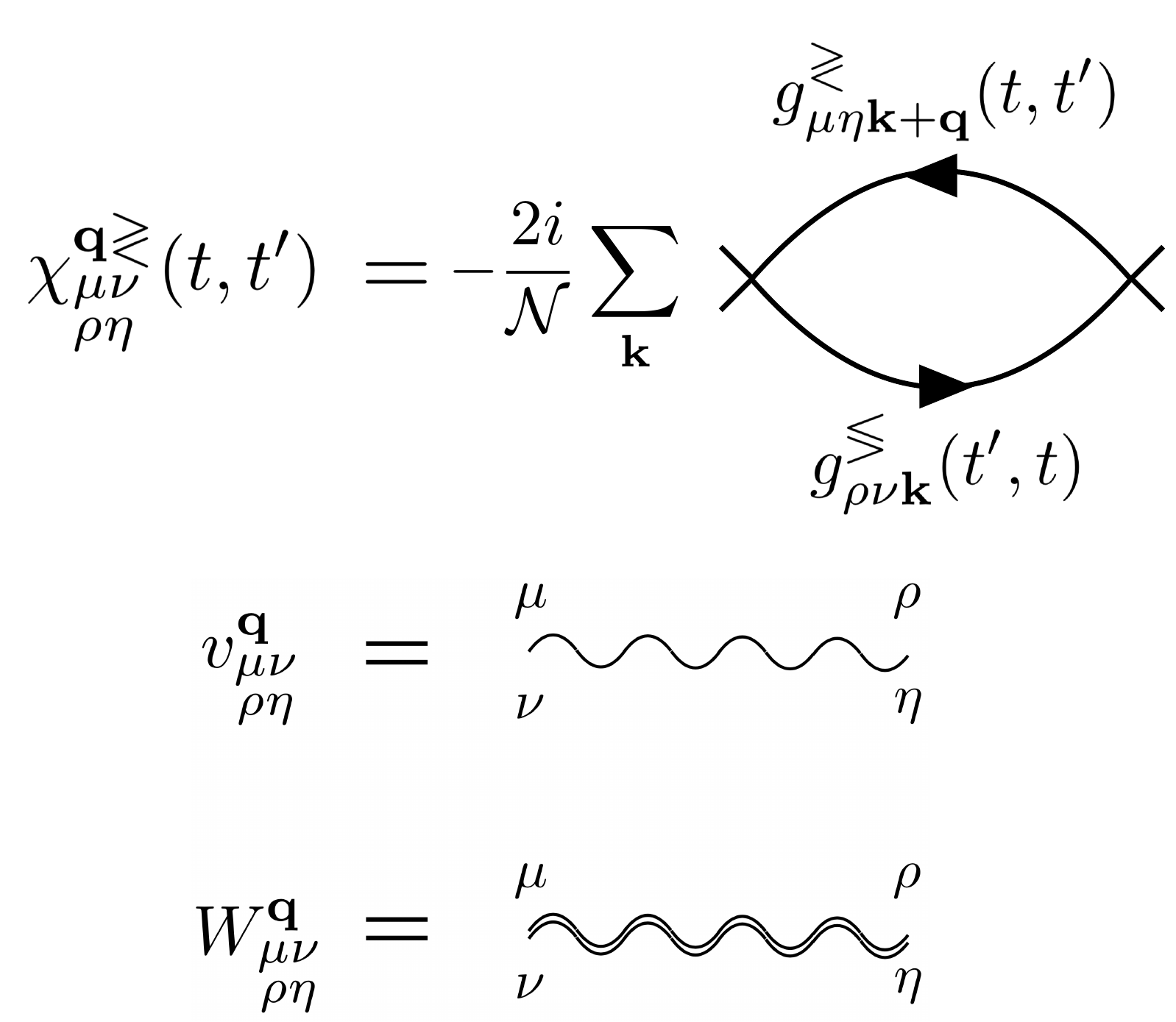}
\caption{Index-convention for the polarization $\chi$ and for 
the bare $v$ and screened $W$ Coulomb repulsion.}
\label{polarization}
\end{figure}

In order to calculate the RPA screened interaction 
in the NEQ-EI phase we need to evaluate the irreducible 
retarded polarization
\bea
\chi^{\blq R}_{\subalign{& \n\m \\& \eta\r}}(t,t')   &=& 
 \theta(t-t')[\chi^{\blq >}_{\subalign{& \n\m \\& \eta\r}}(t,t')-
 \chi^{\blq <}_{\subalign{& \n\m \\& \eta\r}}(t,t')],
 \label{chiR}
\eea
where the greater and lesser component are shown in 
Fig.~\ref{polarization} and read
\bea
\chi^{\blq \gtrless}_{\subalign{& \n\m \\& \eta\r}}(t,t') &\equiv& 
-\frac{i}{\callN}\sum_{\blk 
\s} g_{\m \eta \blk +\blq \s  \s}^{\gtrless}(t,t')  g_{\r \n \blk \s  
\s}^{\lessgtr}(t',t)  \nn \\
&=&- \frac{2i}{\callN}\sum_{\blk} g_{\m \eta \blk +\blq }^{\gtrless}(t,t')  
g_{\r \n \blk  }^{\lessgtr}(t',t) .
\eea
The  lesser and greater Green's function $g_{\blk \s \s'}^{\lessgtr}(t,t') 
\equiv\d_{\s \s'}g_{\blk}^{\lessgtr}(t,t')$ can be conveniently 
written as
\be
g_{\a \b \blk}^{\gtrless}(t,t')= \mp i \varphi^{\pm}_{\a \blk}\varphi^{\pm}_{\b \blk}
e^{-ie^{\pm}_{\blk}(t-t')} 
 e^{i\frac{\d\m}2(S_\a t -S_\b t')},
\label{gt_AM}
\ee
with $S_v=1$ and $S_c=-1$. In the ground-state band-insulating phase ($\d\m=0$)
the anomalous off-diagonal components vanish and the Green's function 
depends on the time difference $(t-t')$ only. In the NEQ-EI phase 
instead the off-diagnal elements are nonzero (symmetry broken phase) 
and hence the Green's function is no longer invariant under time 
translations. 
Taking into account the explicit form of the Green's function we see 
that  $\chi^{\blq \gtrless}_{\subalign{&\m \n \\& \r \h}}(t,t')$ reads
\begin{align}
\chi^{\blq \gtrless}_{\subalign{& \n\m \\& \eta\r}}(t,t')&=-
\frac{2i}{\callN}\sum_{\blk} 
\varphi^{\pm}_{\m \blk+\blq}\varphi^{\pm}_{\h \blk+\blq}
\varphi^{\mp}_{\r \blk}  \varphi^{\mp}_{\n \blk}
\nn\\ 
&\times e^{-i(e^{\pm}_{\blk+\blq}-e^{\mp}_{\blk})(t-t')}
e^{i\frac{\d\m}2[(S_\m-S_\n)t -(S_\h-S_\r) t']}.
\label{chi_gl_AM}
\end{align}

The NEQ polarization 
has a complex time dependence and therefore the RPA screened 
interaction is, in general, not invariant under time-translations.
For a bare interaction $v^{\blq}_{\subalign{&\m \n \\& \h\r}}(t,t')=
\d(t-t')v^{\blq}_{\subalign{&\m \n \\& \h\r}}$ and screened interaction 
$W^{\blq}_{\subalign{&\m \n \\& \h\r}}(t,t')$  as in Fig.~\ref{polarization}, the RPA equation reads
\be
W^{\blq}_{\subalign{&\m \n \\& \h\r}}(t,t')=\d(t-t') 
v^{\blq}_{\subalign{&\m \n \\& \h\r}}+\sum_{\subalign{&\a 
\b \\& \g \d}} \int \!\!d \bar{t} \,  v^{\blq}_{\subalign{&\m 
\n \\& \b \a}} \chi^{\blq R}_{\subalign{& \b\a \\& \g \d}}(t,\bar{t}) 
W^{\blq}_{\subalign{&\g \d \\& \h\r}}(\bar{t},t) .
\label{dystgeneral}
\ee
In our case, however, the bare interaction couples only
pairs of indices belonging to the same band, i.e., $v^{\blq}_{\subalign{&\m 
\n \\& \r \eta}}=\d_{\m \n}\d_{\r \eta}U^{\blq}_{\m\r}$. Consequently, 
Eq.~(\ref{dystgeneral}) is solved by $W^{\blq}_{\subalign{&\m 
\n \\& \r \eta}}=\d_{\m\n}\d_{\r\h}W^{\blq}_{\m\r}$ with
\be
W^{\blq}_{\m \r}(t,t')=U^{\blq}_{\m\r}\d(t-t') +\sum_{\b \g } 
U^{\blq}_{\m\b}\int d \bar{t} 
\chi^{\blq R}_{\subalign{&\b \b \\& \g \g}}(t,\bar{t})  W^{\blq}_{\g
\r}(\bar{t},t).
\label{AM_dyst}
\ee
Taking into account Eq.~(\ref{chi_gl_AM}) we see that $\chi^{\blq 
R}_{\subalign{&\b \b \\& \g \g}}(t,\bar{t})$, defined in 
Eq.~(\ref{chiR}), depends only on 
$t-\bar{t}$. Its explicit form is given below
\begin{align}
\chi^{\blq R}_{\subalign{&\b\b  \\& \g \g}}(t,\bar{t})&= -i\theta(t-t')
\frac{2}{\callN}\nn\\
&\times \sum_{\blk}  \left[  
 \varphi^{+}_{\b \blk +\blq}  \varphi^{+}_{\g \blk +\blq} 
 \varphi^{-}_{\g \blk }  \varphi^{-}_{\b \blk} e^{-i(e^{+}_{ \blk 
 +\blq}-e^{-}_{ \blk })(t-\bar{t})} \right.
 \nn\\
 &\left.
 \varphi^{-}_{\b \blk +\blq}  \varphi^{-}_{\g \blk +\blq} 
 \varphi^{+}_{\g \blk }  \varphi^{+}_{\b \blk} e^{-i(e^{-}_{ \blk 
 +\blq}-e^{+}_{ \blk })(t-\bar{t})} 
\right].
\end{align}
This has an important  consequence since the Fourier transform of 
Eq.~(\ref{AM_dyst}) becomes simply
\be
W^{\blq}_{\m \r}(\w)= U^{\blq}_{\m\r}
+\sum_{\b \g } U^{\blq}_{\m\b}\chi^{\blq R}_{\subalign{&\b \b \\& \g \g}}(\w) 
W^{\blq}_{\g
\r}(\w) ,
\label{dysw}
\ee
where
\begin{align}
\chi^{\blq R}_{\subalign{&\a \a \\& \b \b}}(\w) 
=\frac{2}{\callN}\sum_{\blk} \left[
\frac{ \varphi^{+}_{\a \blk +\blq}  \varphi^{+}_{\b \blk +\blq} 
 \varphi^{-}_{\a \blk }  \varphi^{-}_{\b \blk}}{\w-(e^{+}_{ \blk 
 +\blq}-e^{-}_{ \blk })+i\eta}\right.
 \nn\\
 \left.-
 \frac{ \varphi^{-}_{\a \blk +\blq}  \varphi^{-}_{\b \blk +\blq} 
 \varphi^{+}_{\a \blk }  \varphi^{+}_{\b \blk}}{\w-(e^{-}_{ \blk 
 +\blq}-e^{+}_{ \blk })+i\eta}
\right].
\label{chiw}
\end{align}

More analytic manipulations are possible by taking  
the inter-band repulsion identical to the intra-band repulsion, i.e.,
$U^{\blq}_{\m\n}=U^{\blq}$ (in this case 
the Hartree contributions in $\d V^{\rm HF}_{\blk}$
cancel out). In fact, Eq.~(\ref{dysw}) is then solved by 
$W^{\blq}_{\m \r}(\w)=W^{\blq}_{\rm exciton}(\w)$ with {\it exciton 
screening}
\be
W^{\blq}_{\mathrm{exciton}}(\w)=\frac{U^{\blq}}{1-U^{\blq}
\left[\chi^{\blq R}_{\subalign{&vv \\& vv}}(\w) +\chi^{ \blq R}_{\subalign{&cc 
\\& cc}}(\w) +2\chi^{\blq R}_{\subalign{&vv \\& cc}} (\w)\right]},
\label{hsex}
\ee
where we have used the symmetry property $\chi^{\blq R}_{\subalign{&vv \\& cc}}=
\chi^{\blq R}_{\subalign{&cc \\& vv}}$.
In the following we consider only the {\it static} screening and 
calculate all quantities  at $\w=0$: 
$W^{\blq}_{\mathrm{exciton}}(0)\equiv W^{\blq}_{\mathrm{exciton}}$ and 
$\chi^{\blq R}(0)\equiv\chi^{\blq R}$.

It is worth comparing the 
screening in the NEQ-EI phase with the screening of the 
unbroken symmetry phase ($\D=0$). In this phase the system
can either be a band insulator
or a normal metal, depending on the value of $\d\mu$. In both cases 
the anomalous components of the Green's 
function vanish, hence  $\chi^{\blq R}_{\subalign{&vv \\& cc}}=0$, 
and the screened interaction 
reduces to
\be
W^{\blq}_{\mathrm{plasma}}=\frac{U^{\blq}}{1-U^{\blq}
(\chi^{\blq R}_{\subalign{&vv \\& vv}} +\chi^{\blq R}_{\subalign{&cc 
\\& cc}} )},
\label{wnorm}
\ee
where 
\be
\chi^{\blq R}_{\subalign{&\a \a \\& \a\a}} =\frac{2}{\callN} 
\sum_{\blk}\frac{\bar{f}_{\a \blk+\blq} f_{\a \blk}-f_{\a \blk+\blq} 
\bar{f}_{\a \blk}}{-\e^{\rm HF}_{\a \blk+\blq}+\e^{\rm HF}_{\a \blk}+i\eta} 
\label{normscr}
\ee
is the Lindhard function of a noninteracting gas made of conduction 
electrons up to energy $\m_{c}$ and valence electrons up 
to energy $\m_{v}$ (we remind that  $f_{\a \blk}=0$ for $\e^{\rm 
HF}_{\a 
\blk}>\m_{\a}$ and unity otherwise).
Clearly Eq.~(\ref{normscr}) is non zero only in the metallic case 
and we recover the {\it plasma screening}  of metals for which
the interaction
is maximally screened at $\blq=\mathbf{0}$.
Indeed  the static polarization of an electron gas 
is  real  and negative, reaching its minimum value for  $\blq \to 
\mathbf{0}$. 
% interaction is simply due the sum of two
% standard metallic contributes ({\it plasma screening}), generated separately in the valence and 
% conduction bands in the presence of different chemical potentials $\m_{v}$ and $\m_{c}$.
We also observe that the plasma screening correctly vanishes in the 
band insulating phase ($0<\d\m<\e_{\rm x}$) 
since 
 $\chi^{\blq R}_{\subalign{&vv \\& vv}} = 
 \chi^{\blq R}_{\subalign{&cc \\& cc}}=0$, leading to 
$W^{\blq}_{\mathrm{plasma}}=U^{\blq}$.

The exciton screening in Eq.~(\ref{hsex}) is qualitatively and 
quantitatively different.
Since we have chosen real and normalized eigenvectors we always have
\be
\varphi^{-}_{c\blk}=\varphi^{+}_{v\blk} \quad,\quad
\varphi^{+}_{c\blk}=-\varphi^{-}_{v\blk},
\label{symm}
\ee
and therefore the long-wavelenght limit  $\blq \to \mathbf{0}$ of 
Eq.~(\ref{chiw})  yields  
\be
\chi^{R \blq=\mathbf{0}} _{\subalign{&vv \\& vv}}
=\chi^{R \blq=\mathbf{0}} _{\subalign{&cc \\& cc}} 
=-\chi^{R \blq=\mathbf{0}}_{\subalign{&vv \\& cc}}.
\label{cancellation}
\ee
Consequently the denominator of the screened interaction in 
Eq.~(\ref{hsex}) is unity and
\be
W^{\blq=\mathbf{0}}_{\mathrm{exciton}}=U^{\blq=\mathbf{0}}.
\ee
% We first show some important symmetries of the static 
% polarization tensor in the long-wavelenght limit  $\blq \to \mathbf{0}$.
% Since $V^{cv}_{\blq}=V^{vc}_{\blq}$
% we can always choose
% 
% By using these symmetry relations together with Eq.~(\ref{chiw}) we 
% get the remarkable property
% As a consequence in each entry of the matrix in Eq.~(\ref{4mat}) the anomalous components of $\chi$
% cancel exactly the normal ones, leading to 
% $W^{\blq=\mathbf{0}}_{\mathrm{exciton}}=U_{\blq=\mathbf{0}}$. 
This is a remarkable property conveying a clear physical message: in 
the NEQ-EI phase the long-wavelength limit of the interaction 
is not screened. With hindsight we may say that in a NEQ-EI electrons 
and holes pair to form bound excitons which
behave like microscopic electric dipoles; hence their screening 
efficiency at long distances is correctly 
negligible. 
% This property delivers us the important result that  {\it when exciton condensation occurs the 
% long-wavelength component of the static interaction is not screened}. 
% This finding has also a clear physical interpretation: In the 
% NEQ-EI phase the holes created in the valence band and the electrons 
% promoted in the conduction band form bound excitons, which are 
% neutral composite quasi-particles; at long distances, the screening 
% efficiency of such a gas of microscopic electric dipoles is correctly 
% negligible. 
Notice that starting from the NEQ-EI phase and 
reducing $\d\m$ it is not possible to recover the plasmon screening 
of the unbroken symmetry phase
since the limit $\d\m\to 0$ and $\d\blq\to 0 $ do not commute, see 
also Fig.~\ref{chi}. We also observe that the perfect cancellation 
in Eq.~(\ref{cancellation}) is a consequence of the assumptions 
made, i.e., two-band model and a Coulomb tensor independent of the 
band indices. However this analytic result points to a strong 
cancellation between the anomalous and normal polarization in more 
refined descriptions, and hence to a considerably reduced screening in 
the NEQ-EI phase as compared to that of quasi-free carriers.

\section{Hartree plus Screened Exchange NEQ-EI}
\label{sec3}

In this Section we describe the self-consistent procedure to study  
the NEQ-EI phase within the HSEX approximation.
The equation to be solved is the same as the HF one, see 
Eq.~(\ref{scpdeltarho}), except that  we have 
to replace $U_{\blq} \to W^{\blq}_{\mathrm{exciton}}$ in the {\it exchange} terms, i.e.
\begin{widetext}
\be 
\left(
\begin{array}{cc}
  \e^{\rm HF}_{v \blk}+  \frac{1}{\callN} \sum_{\blq} W^{\blk-\blq}_{\mathrm{exciton}}
 |\varphi^{-}_{c \blq}|^{2}  -\frac{\d\m}{2}&  - \frac{1}{\callN} \sum_{\blq} 
    W^{\blk-\blq}_{\mathrm{exciton}}
  \varphi^{-}_{c \blq} \varphi^{-}_{v \blq}  \\ 
    - \frac{1}{\callN} \sum_{\blq} 
    W^{\blk-\blq}_{\mathrm{exciton}}
  \varphi^{-}_{c \blq} \varphi^{-}_{v \blq} & \e^{\rm HF}_{c \blk}- \frac{1}{\callN} \sum_{\blq} 
    W^{\blk-\blq}_{\mathrm{exciton}}
 |\varphi^{-}_{c \blq}|^{2}  +\frac{\d\m}{2}  \\ 
\end{array}
\right)  \left(
\begin{array}{c}
    \varphi^{\l}_{v \blk}  \\ \varphi^{\l}_{c \blk} 
\end{array}
\right) =   (e^{\l}_{\blk}+\m) \left(
\begin{array}{c}
     \varphi^{\l}_{v \blk}  \\ \varphi^{\l}_{c \blk} \end{array}
\right) .
\label{hhsex}
\ee
\end{widetext}

A working algorithm to solve the problem is proposed below:

\begin{enumerate}
    \item  Solve the HF problem, i.e., solve self-cosistently 
    Eq.~(\ref{scpdeltarho});\\
    \item  Use $\varphi^{\l}_{\m \blq}$ and $e^{\l}_{\blq}$ to calculate $W^{\blq}_{\mathrm{exciton}}$
    %$W^{\blq}_{\subalign{&\m\m \\& \n 
    %\n}}$
    in   Eq.~(\ref{hsex});\\ 
    \item  Use $W^{\blq}_{\mathrm{exciton}}$ %$W^{\blq}_{\subalign{&\m\m \\& \n \n}}$ 
    to  solve self-cosistently the HSEX problem  
    in Eq.~(\ref{hhsex});\\ 
    \item  Use the new $\varphi^{\l}_{\m \blq}$ and $e^{\l}_{\blq}$ 
    to update $W^{\blq}_{\mathrm{exciton}}$ %$W^{\blq}_{\subalign{&\m\m \\& \n \n}}$
    in Eq.~(\ref{hsex});
    \item Repeat steps 3 and 4 until convergence. 
\end{enumerate}
In the next Section we discuss the results of the above 
numerical scheme
for  a 2-dimensional (2D) semiconductor with 
parabolic band dispersion.

\section{Two-Dimensional model for $\mathrm{MoS}_{2}$}
\label{mos2}

We consider a $\mathrm{MoS}_{2}$ monolayer and approximate the valence 
and conduction bands close to the K and K' valleys with the parabolic 
dispersion   $\e^{\rm HF}_{v \blk}=-\frac{k^{2}}{2m}-\frac{\e_{g}}{2}$ and 
$\e^{\rm HF}_{c \blk}=\frac{k^{2}}{2m}+\frac{\e_{g}}{2}$ respectively, with 
$k=|\blk |$. We can 
easily determine the value of the chemical potential $\m$ to fulfill 
the charge neutrality condition in Eq.~(\ref{cncdr}). 
Since $\e^{\rm HF}_{c \blk}=-\e^{\rm HF}_{v \blk}$ and $U^{\blq}_{\m\n}=U^{\blq}$
we have $\m=0$ for all $\d\m$.
According to Ref.~\cite{excscr4} an accurate 
parametrization of the Coulomb interaction $U^{\blq}$ for particles 
in one of the two valleys  is 
\be
U^{\blq}=\frac{V^{\blq}}{\e_{\blq}} ,
\ee
where
\bea
 V^{\blq}&=&\frac{2\p}{q(1+\g q + \d q^{2})}, \nn \\
 \e_{\blq}&=&\e^{\infty}_{\blq}\frac{1-\b_{1 
 \blq}\b_{2\blq}e^{-2hq}}{1+(\b_{1 \blq}+\b_{2 \blq})e^{-hq}+\b_{1 
 \blq}\b_{2\blq}e^{-2hq}}, \nn \\
 \b_{i \blq} 
 &=&\frac{\e^{\infty}_{\blq}-\e_{\mathrm{sub,i}}}{\e^{\infty}_{\blq}+\e_{\mathrm{sub,i}}}, \nn \\
 \e^{\infty}_{\blq} &=& g+\frac{a+q^{2}}{\frac{a\sin(qc)}{qbc}+q^{2}}.
\eea
The dielectric constant $\e^{\infty}_{\blq}$ accounts for the 
 background screening originating from the electronic bands which  are 
neglected, while $\e_{\mathrm{sub,i}}$ is the dielectric constant of a 
possible substrate ($i=1$) or superstrate ($i=2$). 
Realistic parameters to describe a free-standing layer of MoS$_{2}$ 
are $m/m_{e}=0.6$ (where $m_{e}$ is the free electron mass), 
$\e_{g}=2.72~\mathrm{eV}$, $a=2.3~\AA^{-2}$, $b=17$, $c=5~\AA$, $h=2.7~\AA$, $g=5.7$, 
$\g=1.9~\AA$, $\d=0.395~\AA^{2}$, and $\e_{\mathrm{sub,i}}=1$.
We have verified that the solution of the Bethe-Salpeter equation with
the above parameters provides the lowest excitonic level at
$\e_{\mathrm{x}}=2.15~\mathrm{eV}$ with corresponding binding energy 
$\e_{b}=0.57~\mathrm{eV}$, in good agreement with the 
literature~\cite{plasma1}.

\begin{figure}[tbp]
\includegraphics[width=0.4\textwidth]{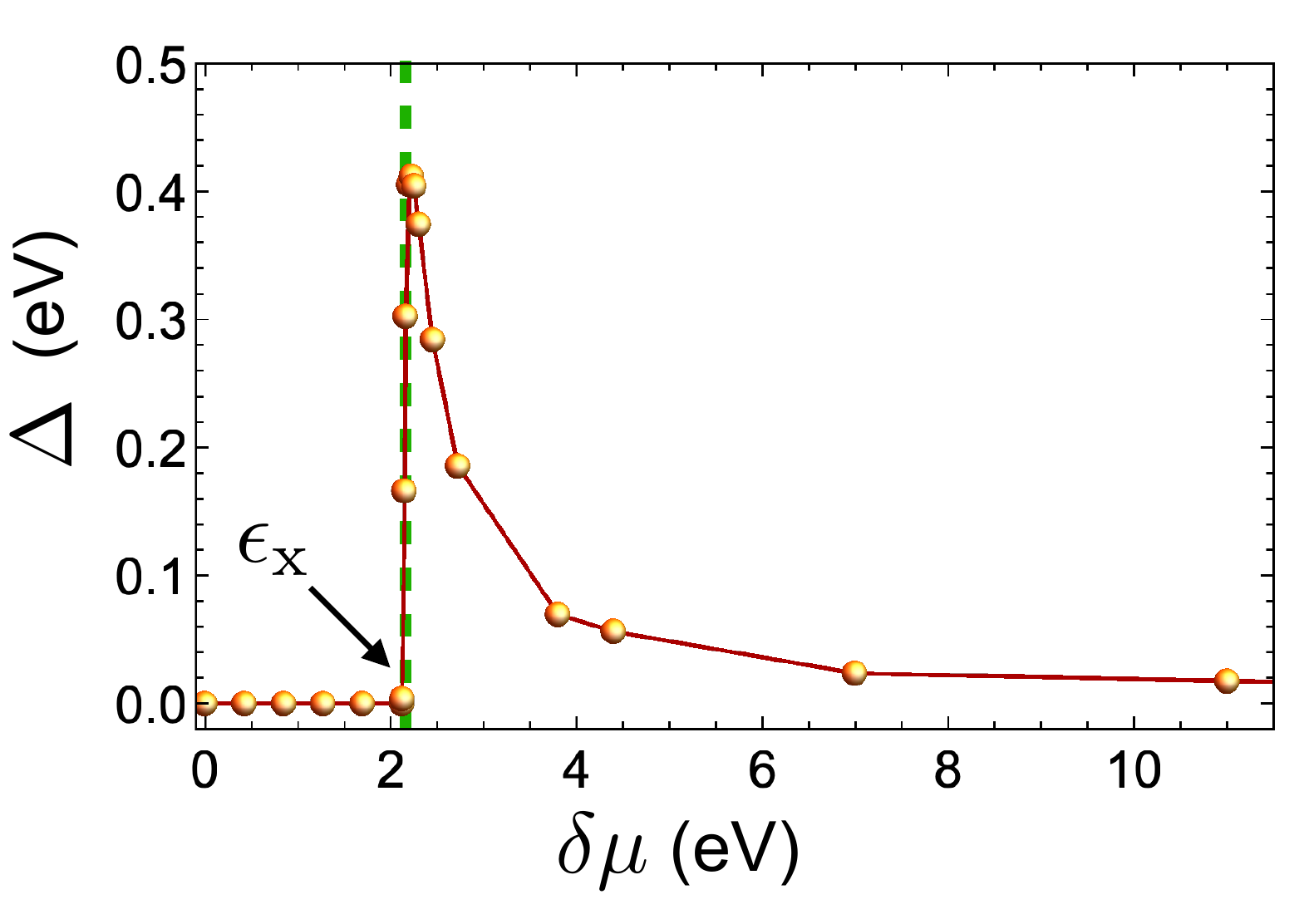}
\caption{HF phase diagram. Order parameter $\D$ defined 
in Eq.~(\ref{orpar}) obtained from the self-consitent solution of 
Eq.~(\ref{scpdeltarho}) for different values of $\d \m$. The vertical 
dashed line indicates the lowest exciton energy 
$\e_{\mathrm{x}}=2.15~\mathrm{eV}$.}
\label{fighf}
\end{figure} 

\subsection{HF phase diagram}

In Fig.~\ref{fighf} we show the NEQ-EI phase diagram 
by displaying the amplitude of the order parameter $\D$ defined in
Eq.~\ref{orpar} versus $\d \m$. 
As discussed in Ref.~\onlinecite{neqei11} 
the order parameter $\D$ and the excited density in conduction band 
$n_{c}=\frac{2}{\callN}\sum_{\blk}|\varphi^{-}_{c 
\blk}|^{2}$ vanish  for $\d \m < \e_{\mathrm{x}}$. 
The transition between the band-insulating and the NEQ-EI phases occurs
at the critical value  $\d \m=\e_{\mathrm{x}}=2.15$~eV.
As $\d \m$ is increased, $\D$ displays a non monotonous behavior characterized by a sudden 
raise  followed by a slow decrease. We checked that $\D$ is not 
discontinuous in $\d \m=\e_{\mathrm{x}}$ and that it reaches  
its maximum value at  $\d \m \gtrsim 2.22$~eV.

\subsection{HSEX phase diagram}

\begin{figure}[tbp]
\includegraphics[width=0.4\textwidth]{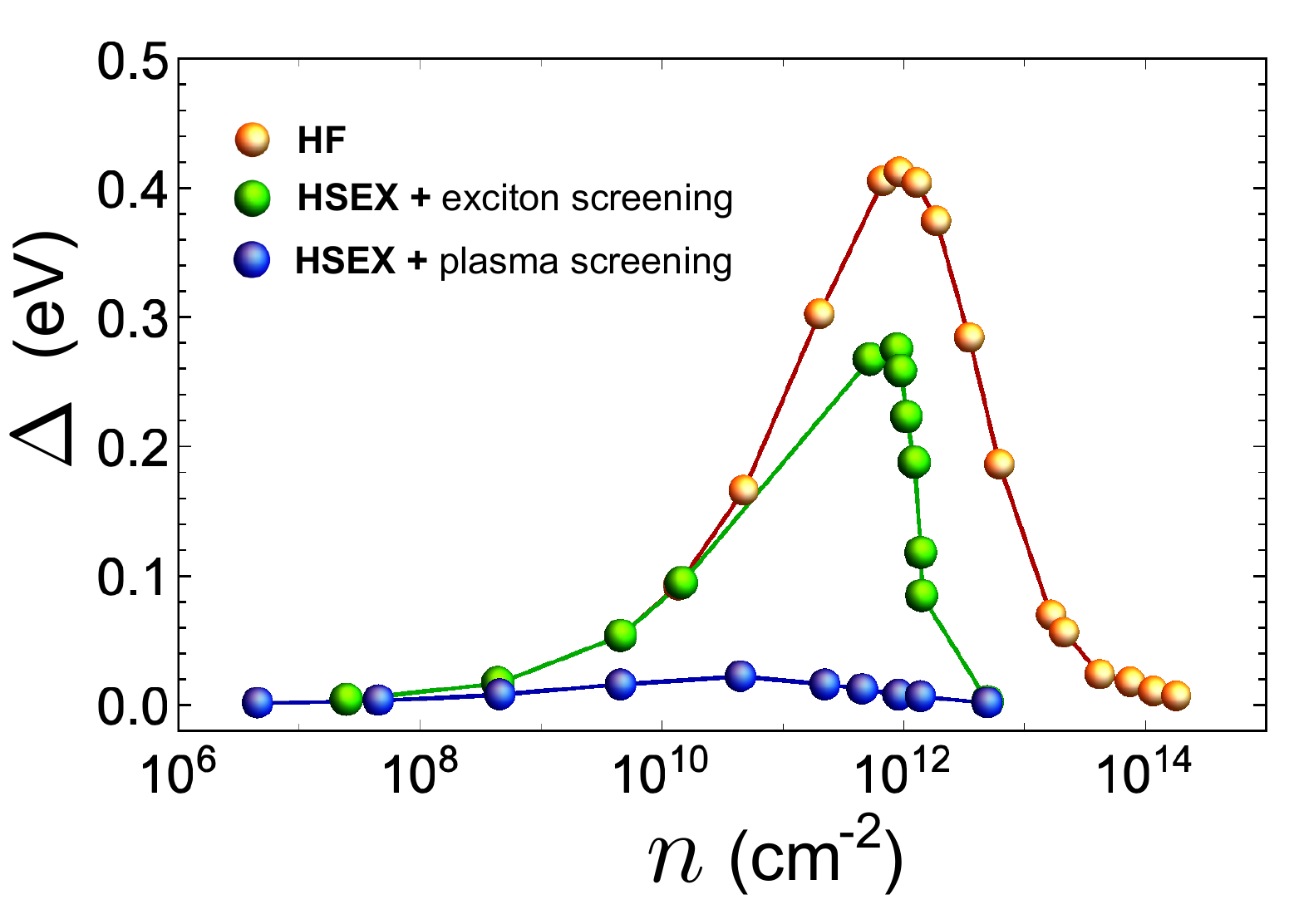}
\caption{HSEX vs HF phase diagrams. Order parameter 
$\D$ calculated in the HF approximation according to 
Eq.~(\ref{scpdeltarho}) (yellow circles) 
and in the HSEX approximation according to Eq.~(\ref{hhsex}) (green circles)
as a function of the excited denisty $n$.
For comparison we also show $\D$ obtained using a plasma screening (blue circles), 
i.e. by solving Eq.~(\ref{hhsex}) with the replacement 
$W^{\blq}_{\mathrm{exciton}} \to W^{\blq}_{\mathrm{plasma}}$.} 
\label{fighsex}
\end{figure} 

In Fig.~\ref{fighsex} we compare the HF phase diagram 
with the HSEX one. For this comparison we have found more instructive to
plot the order parameter $\D$ versus the excited density 
per unit cell, i.e., $n=2n_{c}/A$, where $A=8.8\times 10^{-16}~\mathrm{cm}^{2}$
is the area of the unit cell of  
a $\mathrm{MoS}_{2}$ monolayer. The extra factor $2$ accounts for the fact that 
the total excited carriers are equally distributed among the $\mathrm{K}$ and 
$\mathrm{K}'$ valleys. 
\begin{figure}[tbp]
   \includegraphics[width=0.4\textwidth]{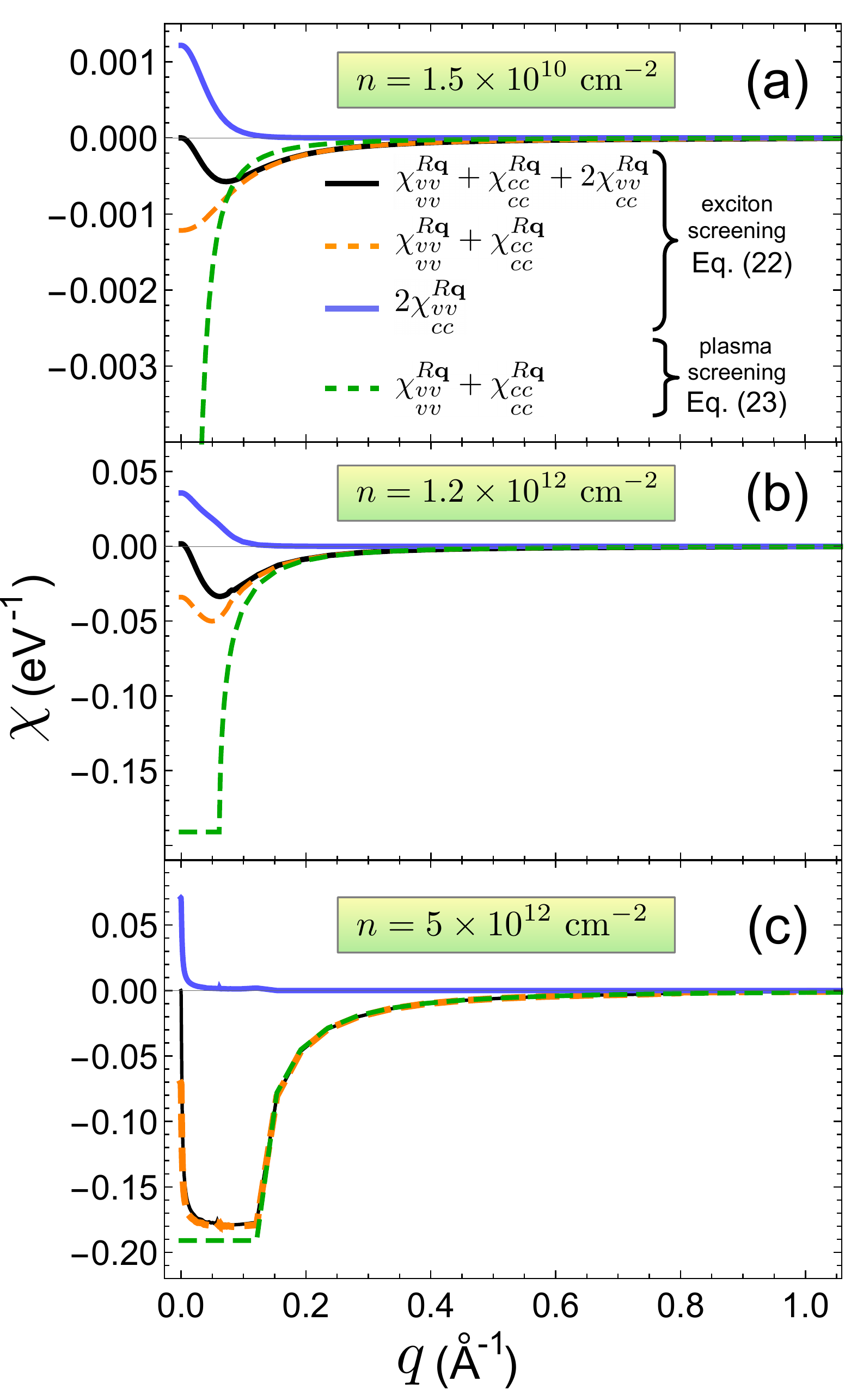}
   \caption{
   The relevant components of the polarization $\chi$ for
   different excited densities $n$ and for different screening 
   approximations.
   In the case of excitonic screening the total $\chi$ (black solid curve)
   entering in Eq.~(\ref{hsex}) has both normal (orange dashed curve) 
   and anomalous (blue solid curve) components.
   In the case of plasma screening, instead, the total $\chi$ entering in 
   Eq.~(\ref{wnorm}) coincides with its 
   normal component (green dashed curve).
   }
   \label{chi}
 \end{figure}
In addition to the full HSEX solution 
we also show the outcome of the 
self-consistent solution of Eq.~(\ref{hhsex}) 
with the replacement $W^{\blq}_{\mathrm{exciton}} \to 
W^{\blq}_{\mathrm{plasma}}$ in Eq.~(\ref{wnorm}).
This comparison is useful to highlight the importance
of screening the interaction with a polarization 
originating from an exciton superfluid rather than from  a plasma 
of free carries.
For parabolic 2D bands the polarization in Eq.~(\ref{normscr}) has an analytic 
expression~\cite{vignale}:
\be
\chi^{R \blq}_{\subalign{&vv  \\& vv}}=\chi^{R \blq}_{\subalign{&cc  
\\& 
cc}}=-\frac{m}{\pi}\left[1-\theta(q-2k_{\mathrm{F}})\sqrt{1-4k_{\mathrm{F}}^{2}/q^{2}}\right],
\ee
where $k_{\mathrm{F}}=\sqrt{2\pi n_{c}}$.

Figure~\ref{fighsex}  clearly shows that the NEQ-EI phase survives 
in a large portion of the phase diagram provided that the proper screened 
interaction is considered. In particular for low and moderate excited densities $n\lesssim 
10^{11}\mathrm{cm}^{-2}$ the screening efficiency of the exciton 
superfluid is  scarce and the HF and HSEX results are quite 
similar. In this regime, however, the plasma screening is already
strong and the corresponding order parameter is highly suppressed.
The dramatic impact of different screenings  on the phase 
diagram  can be better understood with the help of Fig.~\ref{chi}a. For 
low excited density $n$ the excitonic polarization  is much 
smaller than the plasma one. In 
particular the plasma $\chi$ (green dashed) is large 
and independent on $n$ for small 
$\blq$ (its value at $\blq=\mathbf{0}$ is $-m/\p \approx 
-0.19~\mathrm{eV}^{-1}$) whereas the excitonic $\chi$ (black solid) is 
almost vanishing due to the cancellation between
off-diagonal (blue solid) and
diagonal  (orange dashed) components, see  discussion in Section~\ref{screensec}. 
For higher densities the screening of the exciton superfluid becomes more efficient
although the normal and anomalous 
components of $\chi$ still partially cancel at low momenta $\blq$,  
see Fig.~\ref{chi}b.
As a result the HSEX order parameter is somewhat reduced and it reaches
its maximum value concomitantly with the HF order parameter at 
density $n_{\mathrm{max}}\approx 10^{12}\mathrm{cm}^{-2}$. 
For this density the screening is responsible for a 25\% 
reduction of the HF order 
parameter. At the same density the plasma 
screening does instead  suppress the order parameter by two orders of 
magnitude. 
For $n>n_{\mathrm{max}}$ the HSEX results depart 
significantly from the HF values, see Fig.~\ref{fighsex}. 
In particular the HF order parameter 
decreases smoothly whereas in  HSEX 
$n=n_{\mathrm{max}}$ is, {\em de facto}, a critical value beyond 
which the NEQ-EI phase breaks down. 
We refer to this density-driven transition as {\it coherent excitonic Mott 
transition}. This should not be confused with the well-known
excitonic Mott transition~\cite{mott1,mott2,mott3,mott4} which, instead,
refers to the incoherent regime.

The observed behavior can again be understood by inspecting the 
polarization, see Fig.~\ref{chi}c.
At densities $n \gtrsim n_{\mathrm{max}}$ the excitons start  melting
and the screening efficiency changes, becoming similar to the plasma 
efficiency. In fact, although 
$\chi$ still vanishes at $\blq=\mathbf{0}$ (this is an exact property 
for any $\D \neq 0$) the aforementioned cancellation occurs only in a very tiny 
region around $\blq=\mathbf{0}$. 
% Interestingly, a
% density-driven transition occurs in incoherent exciton fluids as well 
% (Mott transition).

The phase diagram in Fig.~\ref{fighsex} provides a reliable description
of MoS$_{2}$ up to $\d \m \lesssim 3.2$~eV. Indeed, in this range 
the  chemical 
potentials $\m_{c}$/$\m_{v}$  lie 
about $0.2$~eV above/below the band miminum/maximum, and the 
parabolic approximation for the band dispersion is still 
accurate~\cite{exc1}.   
The value $\d \m=3.2$~eV corresponds to $n \sim 5\times 
10^{12}\mathrm{cm}^{-2}$, thus covering the whole range of our HSEX 
calculations. 

\section{Summary and Conclusions}
\label{concl}

We presented a microscopic approach to address the stability of the  
exciton superfluid created by a resonant pump 
against an increasing density in the conduction bands.
Using different chemical potential for valence and 
conduction electrons self-consistency naturally leads to the 
non-stationary NEQ-EI state~\cite{neqei11}.
Our theory improves over previous studies in the RPA screened electron-hole interaction 
which we here calculate using the polarization of the 
proper state of matter, i.e., the exciton superfluid. 
We find that the screening does not affect the long-wavelength component
of the  interaction due to the neutral nature of the 
excitons. This  property origins from a 
subtle cancellation between a plasma-like contribution 
and an anomalous one. 
Inclusion of the proper screening in a 
self-consistent HSEX calculation indicates that 
the NEQ-EI phase is very robust, and can survive
up to densities typically excited in  pump-probe experiments.
Numerical calculations in MoS$_{2}$ monolayers show that
the HF (i.e. unscreened) phase diagram is very similar to the 
HSEX phase diagram up to a critical density $n_{\mathrm{max}}\sim 
10^{12}\mathrm{cm}^{-2}$,  where the excitonic order parameter
reaches its maximum value.
However, by further increasing the  
density in the conduction bands excitons start melting consistently with an increase in the 
screening efficiency.
When $n \sim n_{\mathrm{max}}$, the HSEX approach predicts the 
occurrence of a coherent excitonic Mott 
transition.  We 
 do not expect that 
the observed sharpness is universal as other scenarios, 
like phase coexistence, are possible~\cite{capone}.

Our results are relevant also in the light 
of future first-principles studies of the NEQ-EI phase occurring in normal 
semiconductors. Indeed we have provided evidence that at least 
for small and moderate excited densities 
the update of the screened interaction in the 
excited state is presumably not necessary, thus rendering 
the NEQ-EI mean-field problem easily implementable in most of the already 
existing ab initio codes.

 \vspace{1cm}
 
{\it Acknowledgements}  G.S. and E.P. acknowledge funding through
the RISE Co-ExAN (Grant No. GA644076) and the INFN17-nemesys project.
G.S.,  and E.P. acknowledge funding through the MIUR PRIN 
(Grant No. 20173B72NB). A.M., and E.P. acknowledge funding
received from the European Union projects: MaX {\em Materials 
design at the eXascale} H2020-EINFRA-2015-1, Grant agreement n.
676598, and H2020-INFRAEDI-2018-2020/H2020-INFRAEDI-2018-1, 
Grant agreement n. 824143;  {\em Nanoscience Foundries and Fine Analysis - Europe} 
H2020-INFRAIA-2014-2015, Grant agreement n. 654360.
(Grant Agreement No. 654360). G.S. acknowledges Tor Vergata  University
for financial support through the Mission Sustainability Project 2DUTOPI.

%\end{references}

\end{document}

%% file: ourmacros.tex
\newcommand{\be}{\begin{equation}}
\newcommand{\ee}{\end{equation}}
\newcommand{\bea}{\begin{eqnarray}}
\newcommand{\eea}{\end{eqnarray}}

\def\a{\alpha}
\def\b{\beta}
\def\g{\gamma}

\def\d{\delta}
\def\D{\Delta}
\def\e{\epsilon}

\def\h{\eta}

\def\l{\lambda}

\def\m{\mu}
\def\n{\nu}

\def\p{\pi}

\def\r{\rho}
\def\s{\sigma}

\def\vf{\varphi}

\def\w{\omega}

\def\blk{{\mathbf k}}

\def\blq{{\mathbf q}}

\def\callN{\mbox{$\mathcal{N}$}}

\def\Tr{{\rm Tr}}

\def\1op{\hat{\mathbbm{1}}}
\def\nn{\nonumber}
\def\AA{\mathring{\mathrm{A}}}